\documentstyle[epsfig]{mn}

\title[Voids in a universe with CDM and $\Omega_{\Lambda}$]{The formation of
voids in a universe with cold dark matter and a cosmological constant}
\author[O. D. Miranda and J. C. N. de Araujo]
       {Oswaldo D. Miranda$^{1, 2}$\thanks{oswaldo@hbar.wustl.edu} and
       Jos\'e C. N. de Araujo$^2$\thanks{jcarlos@das.inpe.br -- FAPESP Fellow} \\
$^1$Department of Physics, Washington University in St. Louis, Campus Box
1105, One Brookings Drive, St. Louis, MO 63130-4899, USA\\ $^2$Divis\~ao de
Astrof\'\i sica, Instituto Nacional de Pesquisas Espaciais, Av. dos
Astronautas 1758, S\~ao Jos\'e dos Campos, 12227-010 SP, Brazil}

\pagerange{\pageref{firstpage}--\pageref{lastpage}}

\pubyear{2001}

\begin{document}

\maketitle

\label{firstpage}

\begin{abstract}
A spherical Lagrangian hydrodynamical code has been written to
study the formation of cosmological structures in the early
Universe. In this code we take into account the presence of
collisionless non-baryonic cold dark matter (CDM), the
cosmological constant and a series of physical processes present
during and after the recombination era, such as photon drag
resulting from the cosmic background radiation and hydrogen molecular
production. We follow the evolution of the structure since the
recombination era until the present epoch. As an application of
this code we study the formation of voids starting from negative
density perturbations which evolved during and after the
recombination era. We analyse a set of {\it COBE}-normalized models,
using different spectra to see their influence on the formation of
voids. Our results show that large voids with diameters ranging
from $10 h^{-1} {\rm Mpc}$ up to $50 h^{-1} {\rm Mpc}$ can be
formed in a universe model dominated by the cosmological constant
($\Omega_\Lambda \sim 0.8$). This particular scenario is capable
of forming large and deep empty regions (with density contrasts
$\bar \delta < -0.6$). Our results also show that the physical
processes acting on the baryonic matter produce a transition
region where the radius of the dark matter component is greater
than the baryonic void radius. The thickness of this transition
region ranges from about tens of kiloparsecs up to a few
megaparsecs, depending on the spectrum considered. Putative
objects formed near voids and within the transition region would
have a different amount of baryonic/dark matter when compared with
$\Omega_{\rm b}/\Omega_{\rm d}$. If one were to use these galaxies
to determine, by dynamical effects or other techniques, the
quantity of dark matter present in the Universe, the result
obtained would be only local and not representative of the
Universe as a whole.
\end{abstract}

\begin{keywords}
hydrodynamics -- galaxies: formation -- cosmology: observations
-- cosmology: theory -- dark matter -- large-scale structure of
Universe. \end{keywords}

\section{Introduction}

In the 1970s a series of studies showed the existence of
filamentary distribution of galaxies in clusters and superclusters
and large void regions among such filaments. Gregory \& Thompson
(1978) found in galaxy redshift surveys, in the direction of the
Coma cluster, a filamentary structure that is part of the
nowadays denominated `Great Wall' (Geller \& Huchra 1989). This
supercluster is encompassed by a large region of low mass density
(or void).

Additional investigations by Gregory, Thompsom \& Tifft (1981) and
Chincarini, Rood \& Thompson (1981) showed similar results for the
Hercules and Perseus-Pisces superclusters and, later, other
redshift surveys showed that the superclusters are boundaries of
underdense regions (Giovanelli, Haynes \& Chincarini 1986; de
Lapparent, Geller \& Huchra 1986; da Costa et al. 1988).

A redshift survey in the Bo\"{o}tes region (Kirshner et al. 1981)
discovered a large void structure whose estimated diameter is $
\sim 60h^{-1}{\rm Mpc}$. Later, observations showed that this
region in Bo\"{o}tes is not totally empty of matter - it contains
some galaxies (Thuan, Gott \& Schneider 1987; Dey, Strauss \& Huchra 1990).
In spite of this, the void in Bo\"{o}tes has a lower mass
density as compared to the mass density of the Universe.

In particular, Dey et al. (1990) found  21 galaxies
(of which 13 are {\it IRAS} sources) in Bo\"{o}tes, which led the
authors to estimate the average density contrast of this region to
be $-0.84\leq \bar{\delta} \leq -0.66$. More recently, Szomoru
et al. \ (1996a,b) analysed, through the VLA (Very Large Array) technique,
$\sim~1$ per cent of the volume of the void in Bo\"{o}tes ($\sim
1100h^{-3}{\rm Mpc}^3$). These authors found that galaxies in Bo\"{o}tes are
systems rich in gas, composed by late-type galaxies. The study of their
optical properties shows that these systems are very similar to field galaxies
of the same morphological type.

From the theoretical point of view, some models of galaxy
formation (Dekel \& Silk 1986) that studied the cold dark matter (CDM)
scenarios including a bias parameter, predict that dwarf galaxies
should originate from the $1\sigma$ fluctuations, thus being more
smoothly distributed than the rare high-density peaks that form
the more massive galaxies. In these scenarios, the dwarf galaxies
could trace the dark matter and also fill up the voids. On the
other hand, Popescu, Hopp \& Els\"{a}sser (1997), using results of
a search for emission-line galaxies (ELGs), did not find that the
voids are occupied by a homogeneous population of dwarf galaxies.
They found in their sample a few galaxies near the void regions,
but the number of void galaxies is not significant as compared to
the field galaxies.

It is worth stressing that, although the Bo\"{o}tes void has been
one of the first underdense region studied, the distribution of
galaxy clusters present in the Abell catalogue also reveals the
existence of large empty regions with diameters of $\sim
(20-50)h^{-1}{\rm Mpc}$ (Bahcall \& Soneira 1983; Huchra et
al. 1990).

Some authors (Goldwirth, da Costa \& Van de Weygaert 1995) argue that the
typical size of underdense regions (voids) is related to the first
zero of the correlation function $\xi(r)$, which gives $\sim 20
h^{-1} {\rm Mpc}$. This length is less than the void region
sizes found in studies of the large-scale structure of the
Universe.

Recently, El-Ad, Piran \& da Costa (1996), using a particular way to define
voids, have found an average diameter of $37\pm 8h^{-1}{\rm Mpc}$.
In their particular method to define voids they allow the
existence of field galaxies therein, this being a very
reasonable assumption because galaxies could in principle be
formed therein or escape to there, as a result of mutual galaxy
interactions.

Also, quite recently, M\"{u}ller et al. (2000) studied the
distribution of void regions in the two-dimensional slices of the
Las Campanas Redshift Survey (LCRS; the deepest redshift survey
presently available). They found that the distribution of void
sizes scales with the mean galaxy separation, $\lambda$. In
particular, they found that the size of voids covering half of the
area is given  by ${\rm D_{med}}\approx\lambda+(12 \pm 3)= (20-30)
h^{-1}{\rm Mpc}$, where ${\rm D_{med}}$ is the median void
diameter.

In the past, the models addressed to explain the formation of
voids in the distribution of galaxies considered their formation
as a result of the process of clustering (Aarseth \& Saslaw 1982;
Peebles 1982; Vettolani et al. 1985) or as a result of the
evolution, after the recombination era, of negative density
perturbations (Hausman, Olson \& Roth 1983; Hoffman \& Shaham
1983; Filmore \& Goldreich 1984; Bertschinger 1985; de Araujo \&
Opher 1990, 1993).

The formation of voids as a result of
explosions of primordial objects, such as quasars  or other
pre-galactic objets (Ikeuchi, Tomisaka \& Ostriker 1983) was also considered.
This scenario, however, might have some problems because of the distortions
that could be caused on the cosmic background radiation. On the
other hand, this explosive scenario was revisited by Miranda \&
Opher (1996, 1997) and the formation of voids through such a
mechanism may be consistent, within some limits, with the {\it COBE}
satellite data.

In an interesting study, Srianand (1997) has reported the
detection of a large void $\sim 7 h^{-1} {\rm Mpc}$ centred near
the quasi-stellar object (QSO) Tol 1037-2704, where this void could be
produced by an excess of ionization due to the QSO, thereby constituting
another mechanism to produce voids of several megaparsecs.

Also, de Araujo \& Opher (1997) considered an unusual possibility for
the formation of void regions, namely, through putative primordial
magnetic fields.

However, after many years of studies it is not completely clear
how voids are really formed, particularly the largest ones,
such as those found by El-Ad et al. (1996) and M\"{u}ller
et al. (2000). It is possible to have their formation through any of the
mechanisms mentioned above.

It is worth mentioning that, although the dimensions of voids
depend on the particular technique employed to find them (see
e.g. Lindner et al. 1995, 1999;  El-Ad et al. 1996; EL-Ad \& Piran
1997; M\"{u}ller et al. 2000), their very existence
is widely accepted.

Concerning the presence of non-baryonic dark matter in the
structures of the Universe, there is considerable evidence for its
existence. It has been argued that non-baryonic dark matter can
dominate the density and the dynamics of the Universe (for
reviews see Kolb \& Turner 1990; Primack 1997). Thus, there
is no reason to believe that dark matter is not present inside
voids; therefore, it is important to take into account this kind
of matter in the study of the formation of underdense regions, in
order to verify its role.

It is also important to stress that a calculation taking into account the
non-baryonic dark matter and a series of effects during and after the
recombination era, focused on void formation, has so far not been performed.

In the present work we use a spherical Lagrangian hydrodynamical
code, written by one of us (ODM), which allowed us to study the
formation of cosmological structures. Besides the presence of the
non-baryonic dark matter, we consider in our calculations the
expansion of the Universe, the photon drag due to the cosmic
background radiation, the recombination processes and
molecular hydrogen formation. We also included, in our model, the
presence of the cosmological constant to see its influence on the
dimension of void regions.

In Section 2 we describe the model, in Section 3 we present the numerical
results and the discussions, and in Section 4 we present the conclusions.

\section{THE MODEL}

We assume spherically symmetric negative density perturbations
which produce clouds of baryonic and non-baryonic dark matter with
densities lower than the density of the Universe. We treat the
dark matter and the baryonic matter as two fluids coupled by
gravity. We assume, as an initial condition, a top-hat density
profile for the void region.

The hydrodynamic equations that describe the dynamics of the
baryonic density perturbations are

\begin{equation}
{\frac{\partial \rho_{\rm b}}{\partial t}} + {\frac{1}{r^2}}
{\frac{\partial}{
\partial r}}(r^2\rho_{\rm b} v_{\rm b}) = 0,
\end{equation}

\noindent (the mass conservation equation to the baryonic matter
written in spherical coordinates), where $\rho_{\rm b}$ is the
baryonic mass density, $v_{\rm b}$ is the velocity and $r$ is the
radial coordinate.

The equation of motion is given by

\begin{equation}
{\frac{Dv_{\rm b}}{Dt}}=-{\frac 1{\rho_{\rm b}}}{\frac{\partial
P}{\partial r}}-\nabla \Phi - {\frac {4}{3}}{\frac{\sigma_{\rm T}
b T_{\rm r}^4x_{\rm e}}{m_{\rm p} c}}\left[ v_{\rm
b}-H(t)r\right],
\end{equation}

\noindent where $P$ is the pressure of the gas, $x_{\rm e}$ is the
degree of ionization, $b=4\sigma_{\rm SB}/c$ (with $\sigma_{\rm
SB}$ the Stefan-Boltzmann constant), $T_{\rm r}$ is the
temperature of the radiation, $m_{\rm p}$ is the proton mass, $c$
is the velocity of light and $H(t)r$ is the Hubble flow, which
takes into account the expansion of the Universe. The last term
on the right-hand side of equation (2) describes the photon drag and
$D/Dt\equiv \partial /\partial t+v\partial /\partial r$ is the
convective derivative.

The field equation (Poisson's equation) is

\begin{equation}
\nabla^2\Phi = 4\pi G(\rho_{\rm b} + \rho_{\rm d}) - \Lambda\, c^{2},
\end{equation}

\noindent where $\rho_{\rm d}$ is the mass density of the
non-baryonic dark component and $\Lambda$ is the cosmological
constant.

The density parameters of the Universe read

\begin{equation}
\Omega_{\rm b} + \Omega_{\rm d} + \Omega_{\rm k} + \Omega_\Lambda
= 1,
\end{equation}

\noindent where

\begin{equation}
\Omega_\Lambda={\frac{\Lambda\, c^{2}}{3H_0^2}}\;\;\;\;\;\; {\rm and}
\;\;\;\;\;\; \Omega_{\rm k}=-{\frac{k\,c^{2}}{R^2H_0^2}}.
\end{equation}

In the equations (4) and (5), $\Omega_{\rm b}$ ($\Omega_{\rm d}$)
is the baryonic (non-baryonic dark matter) density parameter, $R$
is the scalefactor of the Universe, $H_0$ is the Hubble parameter
(the zero subscript stands for its present value) and $k$ is the
curvature of the Universe, that is, $k=-1$ for an open universe,
$k=0$ for a flat universe and $k=+1$ for a closed universe (thus
$\Omega_{\rm k}$ is the curvature parameter).

The energy equation is written as

\begin{equation}
{\frac{DE}{Dt}} = {\frac{P}{\rho_{\rm b}^2}} {\frac{D\rho_{\rm
b}}{Dt}} - L,
\end{equation}

\noindent where $E$ is the thermal energy per gram and $L$ is the
cooling-heating function of the gas.

The cooling-heating function $L$ is given by the summation of four
mechanisms:

\begin{equation}
L=L_{\rm R} + L_{\rm C} + L_{\rm H2} + L_\alpha.
\end{equation}

(i) The cooling from recombination $L_{\rm R}$ (see, e.g., Schwarz,
McCray \& Stein 1972) is given by:

\begin{equation}
L_{\rm R} = -kN_{\rm A}T_{\rm m}{\frac{Dx_{\rm e}}{Dt}}.
\end{equation}

\noindent The ionization fraction is given by

\begin{equation}
{\frac{Dx_{\rm e}}{Dt}}=C\left[ \beta\, {\rm e}^{-(B_{\rm
1}-B_{\rm 2})/k_{\rm B}T_{\rm r}}(1-x_{\rm e})-{\frac{a\rho_{\rm
b}x_{\rm e}^2}{m_{\rm p}}}\right] +I,
\end{equation}

\noindent where $T_{\rm m}$ is the temperature of the matter,
$N_{\rm A}$ is the Avogadro number, $B_{\rm 1}$ is the bound
energy of the ground state and $B_{\rm 2}$ is the bound energy of the
first excited state. The $C$ and $\beta$ parameters present in equation (9)
are given by

\begin{equation}
C={\frac{\Lambda_{\rm 2s,1s}}{\Lambda_{\rm 2s,1s}+\beta
}}\;\;{\rm and} \;\; \beta =a {\frac{(2\pi m_{\rm e}k_{\rm
B}T_{\rm r})^{3/2}}{h^3}}e^{-B_{\rm 2}/k_{\rm B}T_{\rm r}} .
\end{equation}

\noindent In the above equations $\Lambda_{\rm 2s,1s}=8.227\, {\rm s}^{-1}$ is
the decay rate from the ${\rm 2s}$ state to the ${\rm 1s}$
state, $a=2.84\times 10^{-11}T_{\rm m}^{-1/2}\;\;{\rm cm}^3{\rm
s}^{-1}$ is the recombination rate, $m_{\rm e}$ is the electron
mass and $k_{\rm B}$ is the Boltzmann constant. The collisional ionization
rate $I$ (see, e.g., Defouw 1970) is given by

\begin{equation}
I=1.23\times 10^{-5}N_{\rm A}x_{\rm e}(1+x_{\rm
e})\rho{\frac{k_{\rm B}}{B_{\rm 1}}}T_{\rm m}^{1/2}\, {\rm e}^
{-B_{\rm 1}/k_{\rm B}T_{\rm m}}\;{\rm s}^{-1}.
\end{equation}

(ii) The Compton cooling-heating mechanism $L_{\rm C}$ (
Peebles 1968) is

\begin{equation}
L_{\rm C} = {\frac{4k_{\rm B}N\sigma_{\rm T}bT_{\rm r}^4x_{\rm
e}}{m_{\rm e}c}}(T_{\rm m}-T_{\rm r}).
\end{equation}

(iii) For the cooling by molecular hydrogen $L_{\rm H2}$, we have

\begin{equation}
L_{\rm H2} = {\frac{\Lambda_{\rm M}}{\rho_{\rm b}}}.
\end{equation}

\noindent The equation for $\Lambda_{\rm M}$ and the complete set of
equations for the formation and destruction of ${\rm H_{2}}$
molecules can be obtained in de Araujo \& Opher (1988, 1989) and
Oliveira et al. (1998a).

(iv) The Lyman $\alpha$ cooling $L_\alpha$ (Carlberg 1981) is given by

\begin{equation}
L_\alpha = 1.25\times
10^{-11}C_{12}{\frac{A_{2\gamma}}{A_{2\gamma}+C_{21}}},
\end{equation}

\noindent where $A_{2\gamma }=8.227\;{\rm s}^{-1}$ is the two-photon emission
rate, $C_{21}=1.2\times 10^{-6}T_{\rm m}^{-1/2}x_{\rm e}n_{\rm H}\;{\rm
s}^{-1}$ is the collisional de-excitation rate, $n_{\rm H}$ is the numerical
density of neutral hydrogen and $C_{12}=2C_{21}{\rm e}^{-1.19\times
10^5/T_{\rm m}}$.

The equation of state, to complete the system of equations to the baryonic
matter, is

\begin{equation}
P=k_{\rm B}N_{\rm A}T\rho_{\rm b} (1+x_{\rm e}).
\end{equation}

It is worth stressing that the temperatures in our calculations are never
greater than $\sim 5000{\rm K}$; in this way the cooling and heating mechanisms
considered here account for the thermal history of the baryonic matter
present in the underdense regions that we study.

The equations of conservation of mass and momentum for the dark matter are,
respectively,

\begin{equation}
{\frac{\partial \rho_{\rm d}}{\partial t}} + {\frac{1}{r^2}}
{\frac{\partial}{
\partial r}}(r^2\rho_{\rm d} v_{\rm d}) = 0,
\end{equation}

\begin{equation}
{\frac{Dv_{\rm d}}{Dt}} = - \nabla\Phi.
\end{equation}

We consider that the negative density fluctuations evolve within a
medium of density $\rho_{\rm t}=\rho_0(1+z)^3$, where
$\rho_0=\rho_{\rm bu0}+\rho_{\rm du0}$ (the subscript `u' stands
for the physical parameters of the Universe). Thus, in our model
we take into account the influence of the external medium on the
evolution of the underdense region, and therefore as external
boundary condition we have the pressure of the Universe, namely

\begin{equation}
P_{\rm bu}=k_{\rm B}N_{\rm A}T_{\rm bu}\rho_{\rm bu}(1+x_{\rm
bu}).
\end{equation}

\noindent It is also worth stressing that in many studies of structure
formation the influence of the external medium on the structure under study
is not taken into consideration.

In order to calculate the temperature of the matter of the
Universe we consider in the cooling function only the
contributions due to the Compton cooling-heating and the cooling
due to the recombination processes. The other physical processes
considered for the evolution of the baryonic density perturbation
are not relevant for the evolution of the matter temperature of
the Universe. The energy equation thus reads

\begin{equation}
{\frac{DE_{\rm bu}}{Dt}} = {\frac{P_{\rm bu}}{\rho_{\rm bu}^2}}
{\frac{D\rho_{\rm bu}}{Dt}} - L_{\rm R} - L_{\rm C}.
\end{equation}

Similarly, the collisional ionization is not important for the
matter of the Universe. Thus, we use the study of Peebles (1968),
where the degree of ionization is given by

\begin{equation}
{\frac{Dx_{\rm bu}}{Dt}}=c\left[ \beta\, {\rm
e}^{-(B_1-B_2)/k_{\rm B}T_{\rm r}}(1-x_{\rm bu})-{\frac{
a\rho_{\rm bu}x_{\rm bu}^2}{m_{\rm p}}}\right] ,
\end{equation}

\noindent where now

\begin{equation}
a=2.84\times 10^{-11}T_{\rm bu}^{-1/2}\;\;\;{\rm cm}^3{\rm
s}^{-1}.
\end{equation}

\noindent Equations (19) and (20) are, respectively, the equations of
conservation of thermal energy and ionization balance, with
$T_{\rm bu}$ the temperature of the matter of the Universe and
$x_{\rm bu}$ its ionization fraction.

The baryonic ambient density, i.e., the baryonic density of the Universe,
is given by

\begin{equation}
\rho_{\rm bu}=\rho_{\rm bu0}(1+z)^3={\frac{3H_0^2\Omega_{\rm
b}}{8\pi G}}\left( {\frac{ T_{\rm r}}{T_0}}\right) ^3 ,
\end{equation}

\noindent where $z$ is the redshift and $T_0$ is the present
cosmic background radiation temperature.

The equation for the time evolution of the radiation temperature and,
therefore, the time evolution of the Hubble parameter is given by

%
\begin{eqnarray}
\lefteqn{\left(\frac{\dot{T}_{\rm r}}{T_{\rm r}}\right) = -H(t)
 {} } \nonumber\\ & & {} =-H_0\left[\Omega_{\rm
 m}\left({\frac{T_{\rm r}}{T_0}}\right) ^3 + (1-\Omega_{\rm
 m}-\Omega_\Lambda )\left({\frac{T_{\rm r}}{T_0}}\right) ^2 +
 \Omega_\Lambda \right] ^{1/2},
\lefteqn{ {} } \nonumber\\ & & {}
 \end{eqnarray}
%

\noindent where $\Omega_{\rm m}=\Omega_{\rm b}+\Omega_{\rm d}$ is
the density parameter of the matter.

To solve the above set of equations, a hydrodynamical Lagrangian numerical
code has been written, which basically follows the method developed by
Rychtmyer \& Morton (1967). It is important to stress that the total amount
of baryonic matter and non-baryonic dark matter are taken to be constant
throughout the calculations.

We use in our models 9000 shells for the baryonic matter and
9000 shells for the dark matter. Models with $45\, 000$ shells for
each component showed no considerable difference in the results.

We also test our code, with good agreement, against the one-zone code
of de Araujo \& Opher (1991) for the collapse of  baryonic and non-baryonic
dark matter Population III objects (positive density perturbation). Also, as
an additional test we reproduce the results of de Araujo \& Opher (1993) for
voids formed from negative density perturbations, where a one-zone model was
used. The program conserves mass and total energy of the system to at least
one part in $10^6$.

\section{NUMERICAL RESULTS AND DISCUSSIONS}

In the present paper we use different power spectra of density
perturbations to study their role on the scale of voids
formed and to see if cosmological parameters could be constrained.

First we take a typical CDM power spectrum of perturbations
given by (Padmanabhan 1993)

\begin{equation}
\mid\delta_{\rm k}\mid ^{2} = {\frac{Ak^{(4-2\alpha)}}
{(1+Bk+Ck^{3/2}+Dk^2)^2}},
\end{equation}

\noindent where $A$ is the spectrum amplitude, $B=1.7(\Omega
h^2)^{-1}{\rm Mpc}$, $C=9.0(\Omega h^2)^{-3/2}{\rm Mpc}^{3/2}$ and
$D=1.0 (\Omega h^2)^{-2}{\rm Mpc}^{-2}$. In the above equation the
power index is related to $\alpha$ by the relation
$n=(4-2\alpha)$. The variance of fluctuations on scale $r_0$ reads

\begin{equation}
\Delta^2(r_0)={\frac{1}{2\pi^2}}\int_0^\infty
 dk\,k^2\mid\delta_{\rm k}\mid^2W^2(k\,r_0).
\end{equation}

We analysed two different window functions $W(k\,r_0)$: Gaussian
and top-hat filters. Our results are very weakly dependent on
$W(k\,r_0)$ so, for convenience and to compare our
results with previous studies, we present here the results using
a top-hat window function. Then, we have

\begin{equation}
W_{\rm k} = 4\pi r_0^3 \left[{\frac {\sin (k\, r_0)}{(k\, r_0)^3}}
- {\frac {\cos (k\, r_0)}{(k\, r_0)^2}}\right],
\end{equation}

\noindent with the volume of the top-hat window given by

\begin{equation}
V_{\rm W} = {\frac {4}{3}}\pi r_0^3.
\end{equation}

The mass fluctuation in a radius $R$ reads

\begin{equation}
\left({\frac {\delta \rho}{\rho}}\right)_R^2 = {\frac
{1}{2\pi^2}} \int_0^\infty k^2dk |\delta(k)_{\rm V}|^2 {\frac
{|W_{\rm k}|^2}{V_{\rm W}^2}}.
\end{equation}

\noindent The input parameters that describe the cosmological scenarios
are taken from a recent study performed by Cen (1998a,b), who
analysed mass and autocorrelation functions of rich clusters of
galaxies from linear density fluctuations. Cen presents 32 different
cosmological models to determine the fitting parameters in his Gaussian peak
method, using them to obtain the cluster mass function. In particular, we
take four of these models that present different characteristics related to
the present radii and present density contrasts of the voids. The models
chosen have their parameters described in Table 1 (see also Cen 1998a,b).

\begin{table*}
\caption{The cosmological models chosen from Cen (1998a,b). The
model numbers correspond to those presented in Cen's table 1. We
consider a baryonic component $\Omega_{\rm b}=0.05$ and for the
dark component we have $\Omega_{\rm d}=\Omega_{\rm m}-0.05$.
Models 23 and 26 include the cosmological constant ($\Lambda$).}
\begin{tabular}{@{}lcccccc}
\hline
     Model & $\Omega_{\rm m}$ & $\Omega_\Lambda$ & $\sigma_8$ & $n$ &
     $H_0$ & Comment \\
     & & & & & $({\rm km}\,{\rm s}^{-1}\,{\rm Mpc}^{-1})$ & \\
\hline

14  & $0.35$ & $0$    & $0.80$ & $1.0$  & $70$  & Open CDM  \\
20  & $0.60$ & $0$    & $0.73$ & $1.0$  & $60$  & Open CDM  \\
23  & $0.40$ & $0.60$ & $0.79$ & $0.95$ & $65$  & $\Lambda$ CDM \\
26  & $0.20$ & $0.80$ & $1.5$  & $0.95$ & $100$ & $\Lambda$ CDM \\

\hline
\end{tabular}
\end{table*}

All models considered here have `excess power' $EP=1.30\pm 0.15$.
(`Excess power' is defined as $EP\equiv 3.4 \sigma_{25}/\sigma_8$, where
$\sigma_{25}$ is the linear rms density fluctuation in a $25h^{-1}{\rm Mpc}$
top-hat sphere at $z=0$, and $\sigma_8$ has a similar definition in an
$8h^{-1}{\rm Mpc}$ top-hat sphere.) This
parameter describes the shape of the power spectrum on scales
$\sim 8-300h^{-1}{\rm Mpc}$ and its values range from $1.15$ to
$1.45$ to fit the {\it COBE} data.

In the first column of Table 1 we present the number of the model
[we use here the same numbers presented by Cen (1998a,b)]; in the
second and third columns we present the density parameter for the
matter and that associated with the cosmological constant,
respectively; the fourth column presents the rms density
fluctuation in an $8h^{-1}{\rm Mpc}$ top-hat sphere at $z=0$; in
the fifth column we present the power index $n$; in the sixth
column we present the value of the Hubble constant; and finally the
seventh column indicates the type of power spectrum used.

For the normalization of the power spectra studied here we
followed the usual methodology found in the literature, where the
value of $\sigma_{8}$ permits one to find the constant $A$ of the
power spectra. Our normalized spectra follow the same shape seen
in fig. 2 of Cen (1998a).

Thus, for a given perturbation with mass $M_{\rm C}$, and for the
models described in Table 1, we can obtain the present radius and
the present density contrast. Then, to determine the initial
conditions at the beginning of the recombination era, we used the
general formula given by Opher, Pires \& de Araujo (1997) for the
amplification of the density perturbations (their study
incorporates both growing and decaying modes).

We start the calculations at the beginning of the recombination
era, at redshift $z_{\rm rec} \sim 1000$, where the ionization
degree begins to be significantly lower than unity. For $z_{\rm rec}
> 1000$ the matter is completely ionized and density perturbations
(positive or negative) cannot evolve due to the photon drag that
inhibits the growth of any peculiar velocity. We start our
calculations assuming that the underdense regions evolve initially
with a velocity field given by the Hubble flow.

In the present study we assume that initially

\begin{equation}
\delta_{\rm i}={\frac{\delta \rho_{\rm t}}{\bar{\rho}_{\rm
t}}}={\frac{\delta \rho_{\rm b}}{\bar{ \rho}_{\rm
b}}}={\frac{\delta \rho_{\rm d}}{\bar{\rho}_{\rm d}}},
\end{equation}

\noindent where $\bar{\rho}_{\rm t}$ is the total density (dark
matter and baryons), $\bar{\rho}_{\rm b}$ is the baryon density,
$\bar{\rho}_{\rm d}$ is the dark matter density, and $\delta
\rho_{\rm t}$, $\delta \rho_{\rm b}$ and $\delta \rho_{\rm d}$ are
the respective density perturbations.

The baryonic mass contained within the underdense region is
obtained from

\begin{equation}
M_{\rm b}=M_{\rm C} {\Omega_{\rm b}}/{\Omega_{\rm m}}
\end{equation}

\noindent and, analogously, the collisionless dark matter within
the underdense region is

\begin{equation}
M_{\rm d}=M_{\rm C} {\Omega_{\rm d}}/{\Omega_{\rm m}}.
\end{equation}

The initial radius of the underdense region is given by

\begin{equation}
R_{\rm i}=\left[{\frac 3{4\pi }}{\frac {M_{\rm C}}{\rho_0}}\frac
1{(1+\delta_{\rm i})}\right] ^{1/3}(1+z_{\rm rec})^{-1}.
\end{equation}

\noindent With such an initial profile, the density at the void boundary
does not fall smoothly to the density of the Universe. Thus, the interface
(between the external medium and the last shell of the computational grid)
is a membrane-like one. This membrane is maintained throughout the void
time evolution, because the mass inside the void region is taken to be
constant throughout the void evolution.

In Tables 2-5 we present the main results for the models described
in Table 1. We see that the spectra of perturbations used here
produce different results concerning the final diameter of the
voids (and for their density contrasts) for the same mass
perturbation $M_{\rm C}$.

\begin{table}
\caption{Principal results for model 14 described in Table 1.
The final diameter of the baryonic component is $D_{\rm BF}$,
while $D_{\rm DF}$ is the final diameter of the dark matter
component. The density contrasts for the baryonic and dark matter
components are, respectively, $\delta_{\rm BF}$ and $\delta_{\rm DF}$.}
\begin{tabular}{@{}lcccc}
\hline
     $M_{\rm C} ({\rm M}_\odot)$ & $D_{\rm BF}(h^{-1}{\rm Mpc})$
     & $\delta_{\rm BF}$
     & $D_{\rm DF}(h^{-1}{\rm Mpc})$ & $\delta_{\rm DF}$ \\
\hline

$10^{12}$ & 3.82   & $-0.755$ & 3.89   & $-0.755$ \\
$10^{13}$ & 7.18   & $-0.619$ & 7.28   & $-0.619$ \\
$10^{14}$ & 14.2   & $-0.453$ & 15.7   & $-0.455$ \\

\hline
\end{tabular}
\end{table}

\begin{table}
\caption{Principal results for model 20 described in Table 1.
See Table 2 for notation.}
\begin{tabular}{@{}lcccc}
\hline
     $M_{\rm C} ({\rm M}_\odot)$ & $D_{\rm BF}(h^{-1}{\rm Mpc})$
     & $\delta_{\rm BF}$
     & $D_{\rm DF} (h^{-1}{\rm Mpc})$ & $\delta_{\rm DF}$ \\
\hline

$10^{12}$ & 3.07  & $-0.770$ & 3.11  & $-0.770$  \\
$10^{13}$ & 5.72  & $-0.609$ & 5.80  & $-0.609$  \\
$10^{14}$ & 11.3  & $-0.360$ & 11.5  & $-0.360$  \\

\hline
\end{tabular}
\end{table}

\begin{table}
\caption{Principal results for model 23 described in Table 1.
See Table 2 for notation.}
\begin{tabular}{@{}lcccc}
\hline
     $M_{\rm C} ({\rm M}_\odot)$ & $D_{\rm BF}(h^{-1}{\rm Mpc})$
     & $\delta_{\rm BF}$
     & $D_{\rm DF} (h^{-1}{\rm Mpc})$ & $\delta_{\rm DF}$ \\
\hline

$10^{12}$ & 3.80  & $-0.800$ & 3.90  & $-0.801$ \\
$10^{13}$ & 6.84  & $-0.632$ & 6.93  & $-0.632$ \\
$10^{14}$ & 13.4  & $-0.369$ & 14.2  & $-0.372$ \\

\hline
\end{tabular}
\end{table}

\begin{table}
\caption{Principal results for model 26 described in Table 1.
See Table 2 for notation.}
\begin{tabular}{@{}lcccc}
\hline
     $M_{\rm C} ({\rm M}_\odot)$ & $D_{\rm BF}(h^{-1}{\rm Mpc})$
     & $\delta_{\rm BF}$
     & $D_{\rm DF} (h^{-1}{\rm Mpc})$ & $\delta_{\rm DF}$ \\
\hline

$10^{12}$ & 8.30   & $-0.932$ & 8.40   & $-0.934$ \\
$10^{13}$ & 13.4   & $-0.853$ & 13.5   & $-0.853$ \\
$10^{14}$ & 23.9   & $-0.709$ & 25.0   & $-0.709$ \\
$10^{15}$ & 48.7   & $-0.622$ & 53.5   & $-0.628$ \\

\hline
\end{tabular}
\end{table}

In particular, if we consider that voids are regions with density
contrasts $\bar \delta \leq -0.6$, then large voids with diameters
of $\bar D > 20-30 h^{-1} {\rm Mpc}$ [the sizes found by El-Ad et al.
(1996) and M\"{u}ller et al. (2000)] could be produced only for the
spectrum 26 or similar one (see Tables 1 and 5).

In fact, it is worth mentioning that even if $(2-3)\sigma$ negative
density perturbations are considered for the  models 14, 20 and
23 or similar ones, large void regions cannot be accounted for. For
example, for model 23 (see Tables 1 and 4) a $2\sigma$
negative density perturbation of $10^{14}{\rm M}_{\odot}$ would produce
a void with baryonic (dark matter) diameter of $14.7 h^{-1}{\rm
Mpc}$ ($16.7 h^{-1}{\rm Mpc}$) with $\delta_{\rm BF}=-0.639$
($\delta_{\rm DF}= -0.639$). If instead a $3\sigma$ negative
perturbation is considered, a void with baryonic (dark matter)
diameter of $17.2 h^{-1}{\rm Mpc}$ ($18.9 h^{-1}{\rm Mpc}$) with
$\delta_{\rm BF}=-0.776$ ($\delta_{\rm DF}= -0.776$) would be
produced. Although large, these voids are not enough to explain
those reported in the literature.

It is worth noting that some authors define void regions as having
a density of 0.2 times the mean galaxy density (Schmidt, Ryden
\& Melott 2001). The density contrast in clusters of galaxies is
around 1.5-3.0, which means that the density contrast in void regions
would amount to $ \delta = -(0.5 \pm 0.2)$, which is less
conservative than our definition. Even considering that
regions with $\delta \sim -0.3$ are voids, it is easily seen from
our results that the largest voids in the Universe cannot be
accounted for by models 14, 20 and 23 or similar ones.

A possible explanation for the largest voids could be related to
the interaction and eventual merger of two or more small voids. We
refer the reader to a paper by Dubinski et al. (1993), who
investigated the evolution, interaction and merger of voids formed
from negative density perturbations. Then, the merger of small
voids could be a way to build up large voids from the spectra of
density perturbations related to models 14, 20 and 23.

Still considering our calculations, note that, using model 26 with
$3\sigma$ perturbations for $10^{15}{\rm M}_\odot$, void regions with
diameters of up to $62 h^{-1}{\rm Mpc}$ (baryonic component) with
density contrast $\delta_{\rm BF} = -0.756$ could be produced. In
this case, the dark matter void would have a diameter of $D_{\rm
DF}=69 h^{-1}{\rm Mpc}$ (with $\delta_{\rm DF}= - 0.765$). This result
is roughly that inferred to the diameter of the Bo\"otes void.

An interesting issue has to do with the fact that the
post-recombination Jeans mass is $\sim 10^6 {\rm M}_\odot$ at
recombination and it decreases later on, and that the mass scales
studied here are in the range of $10^{12}{\rm M}_\odot$ to
$10^{15}{\rm M}_\odot$. Thus, one could argue that collapse and
eventual fragmentation should take place in the evolving
(expanding) shells. To address this issue properly one should
study in detail how density perturbations evolve in the expanding
shells.

Note that it is not enough that a perturbation be Jeans unstable
to guarantee that it will collapse and fragment. Consider the
following example. Let us think of the evolution of primordial
clouds after the recombination era (that is, the evolution of
positive density perturbations). Even the clouds that are already
Jeans unstable at the recombination era will initially expand
before collapsing. For a cloud that stops expanding, its mean
density ($\bar \rho_{\rm c}$) relative to the background is given by
(see, e.g., Peebles 1993; Coles \& Lucchin 1995)

\begin{equation}
\left({\bar\rho_{\rm c} \over \rho}\right) = \left({3\pi \over
4}\right)^2\simeq 5.6 ,
\end{equation}

\noindent which corresponds to

\begin{equation}
\left({\delta\rho\over \rho}\right) \simeq 4.6
\end{equation}

This result is a lower limit since it depends only on the
expansion rate of the Universe (the above result is taken for an
Einstein-de Sitter universe) and on the gravity of the cloud. For
the cases in which other physical processes are relevant for the
evolution of the cloud, the turnaround should occurr for

\begin{equation}
\left({\delta\rho\over \rho}\right) > 4.6.
\end{equation}

\noindent In particular, we refer the reader to the papers by
Oliveira et al. (1998a,b) who studied the collapse of
Population III objects and consider such issues in detail.

Concerning the void regions, their density contrast is always
negative and they never stop expanding. As a result the condition
given by equation (35) is never fulfilled and so the evolving
shells do not collapse.

Density perturbations that could be present in the evolving
shells, however, could in principle collapse and fragment if they
had the time to stop expanding and evolve. Note that this issue
concerning the collapse and eventual fragmentation in the evolving
shells is by itself so interesting that it deserves to be
investigated in detail. We leave, therefore, such an issue to be
considered in another paper to appear elsewhere.

An interesting result concerning the distribution of dark matter
and baryonic matter can be seen in Tables 2-5. The final diameter
of the dark matter is always greater than the final diameter of
the baryonic matter, when all physical processes considered in the
present paper are included in the calculations.

To see the influence of the several physical processes on the
evolution of a given negative density perturbation, in particular
on the final diameters of the voids, we performed some
calculations studying individually the effects of all the four
cooling-heating mechanisms and also the influence of the photon
drag (Table 6).

\begin{table*}
\caption {The influence of the physical processes on the evolution
of the final diameter and density contrast of a void formed from a
perturbation with mass $10^{13}{\rm M}_\odot$ in model 23. The
cooling-heating mechanisms for the baryonic matter are: $L_{\rm
R}$ the cooling from recombination, $L_\alpha$ the Lyman $\alpha$
cooling, $L_{\rm C}$ the Compton cooling-heating, and $L_{\rm H2}$
the cooling by molecular hydrogen. In the second and third columns
we present the results without all the physical processes. The
superscript $+{\rm U}$ ($-{\rm U}$) in these columns means that we keep
(remove) the pressure of the Universe acting on the `void boundary'.}

\begin{tabular}{@{}lccccccc}

\hline
& Without all        & Without all        & Without & Without & Without
&Without & Without \\ & processes $^{(+{\rm U})}$ & processes $^{(-{\rm U})}$
&photon drag & $L_{\rm R}$ & $L_\alpha$ & $L_{\rm C}$ &$L_{\rm H2}$ \\
\hline
$D_{\rm BF}(h^{-1}{\rm Mpc})$ & 6.97   & 7.03   &  6.94   & 6.86   & 6.86   & 6.83   & 6.86   \\
$\delta_{\rm BF}$             & $-0.647$ & $-0.648$ & $-0.643$  & $-0.640$ &
$-0.640$ & $-0.639$ & $-0.640$ \\ $D_{\rm DF}(h^{-1}{\rm Mpc})$ & 7.02   &
7.03   &  6.99   & 6.93   & 6.93   & 6.90   & 6.93   \\ $\delta_{\rm DF}$     
       & $-0.648$ & $-0.648$ & $-0.643$  & $-0.641$ & $-0.641$ & $-0.640$ &
$-0.641$ \\ \hline \end{tabular} \end{table*}

Note that the dissipative baryonic processes do not define a
particular characteristic void scale; their main effect is
to segregate the dark matter and baryonic matter.

From Tables 4 and 6 we note that the baryonic diameter is a
little bit lower than the dark matter diameter when all physical
processes are included, and this result remains when we disregard
each one of the cooling-heating mechanisms.

Another interesting result is obtained when we disregard all
the processes. In this case, the final diameter of the baryonic matter
is almost the same as the dark matter component. When we disregard all the
processes and the pressure of the Universe, the final diameters are the
same for both components, as expected, since the two fluids in these
circumstances are subject only to gravity.

In Table 6 we see the influence of each physical process one at a
time. When we disregard the photon drag, but maintain all the
other processes and the pressure of the Universe, the final radius
of the baryonic (dark) component goes from $6.84\; (6.94)\, h^{-1}{\rm
Mpc}$ to $6.93\; (6.99)\, h^{-1}{\rm Mpc}$. This occurs because the photon
drag acts against the expansion of the negative density perturbations,
inhibiting the growth of peculiar expansion velocities.

The cooling mechanisms $L_{\rm R}$, $L_\alpha$ and $L_{\rm H2}$
have very similar effects on the void evolution, and their
principal influence is related to the thermal evolution of the
matter inside the void regions. The Compton heating-cooling when
disregarded reduces slightly the dimension of the void, because at
high redshifts this physical process maintains the temperature
of the matter inside the void close to the radiation temperature.

The above results also appear when we perform the present analysis
for other models described in Table 1. In particular, independent
of the spectra used, the physical processes produce very similar
results concerning the final diameters of the baryonic and dark
matter. We note that even using other spectra, like those used by
de Araujo \& Opher (1993) for example, this difference in
diameter due to the presence of physical processes is maintained.

With all the physical processes included, there exists a
transition region with thickness of up to $\sim 2.5\, h^{-1}{\rm
Mpc}$, which depends on the mass of the perturbation, the
normalization of the spectrum and the respective density
parameters for the dark and baryonic components as well as the
cosmological constant. These transition regions have sizes
comparable to the diameters of galaxies and clusters of galaxies.

The model 26, for example, has a transition region of $\sim
2.5\, h^{-1}{\rm Mpc}$ thickness for a perturbation with mass
$10^{15}{\rm M}_\odot$. Within this region the density
contrast of the cold dark matter is $\delta_{\rm DF}=-0.628$ (its
diameter is $53.5\, h^{-1}{\rm Mpc}$). Thus, the mass density of the dark
matter is $\rho_{\rm d}=0.372\rho_{\rm du}$, where $\rho_{\rm du}$
is the ambient mass density of the dark matter (that is, the mass
density of the dark matter present in the Universe).

As the boundary of the baryonic matter is at $48.7\, h^{-1}{\rm Mpc}$,
within the transition region the density contrast of the baryonic
matter is $\delta_{\rm B}=0 $, and so $\rho_{\rm b}=\rho_{\rm bu}$.
The relation between the baryonic and dark components present within
this transition region is given by

\begin{equation}
\frac{\rho_{\rm b}}{\rho_{\rm d}}=2.70\times \left( {\frac{\rho_
{\rm bu}}{\rho_{\rm du}}} \right) = 2.70\times \left(
{\frac{\Omega_{\rm b}}{\Omega_{\rm d}}}\right) = 0.90
\end{equation}

Thus, in this region the amount of baryonic matter is
approximately equal to the amount of non-baryonic dark matter,
although this universe model has three times more dark matter
($\Omega_{\rm d}=0.15$) than baryonic matter ($\Omega_{\rm b}=0.05$).

Consequently, galaxies formed near voids and within the
transition region could, during their formation process, have a
lower content of cold dark matter than those galaxies formed in
regions far from voids. This result leads us to conclude that, if
one uses an analysis of the content of dark to baryonic matter for
the galaxies in different regions, one could infer very different
results, reflecting therefore only local values.

Let us consider another example. The transition zone of model
26, for a perturbation with mass $10^{13}{\rm M}_\odot$, gives
$\rho_{\rm b} /\rho_{\rm d}  \sim 7 (\Omega_{\rm b}/\Omega_{\rm
d}) \sim 2.3$, although in such a universe model the amount of
dark matter is three times greater than the amount of baryonic
matter.

It is worth stressing that galaxies formed out of the transition
zone and that escape to there could have a different relation
between the baryonic and dark components than that found in the
transition zone.

It is important to have in mind that the process of structure
formation of the Universe is a very complicated one. For the
canonical model of cosmology, namely primordial Gaussian random
perturbation field, the formation of voids and clusters of
galaxies is part of one process of clustering. The formation of
regions with positive density contrasts, like clusters of
galaxies, contributes to the growth of underdense regions and vice
versa.

The present study only considers a small part of the very
complicated structure formation process. We study the evolution of
negative density perturbations taking into account all relevant
physical processes acting on them. However, the assessment of the
role of these very physical processes not considered in the
literature, besides including the presence of non-baryonic dark
matter and cosmological constant, are the main contributions of
the present study.

\section{CONCLUSIONS}

In the present investigation, we study the evolution of negative
density perturbations taking into account a series of physical
processes that are present during and after the recombination era.
In order to perform such a study, a spherical Lagrangian
hydrodynamical code has been written to follow the evolution of
these density perturbations.

We analysed a set of {\it COBE}-normalized spectra obtained from a
recent study performed by Cen (1998a,b). The results presented
here show that negative density perturbations can directly create
large voids, with diameters of up to $\sim 50\, h^{-1}{\rm Mpc}$,
only for some universe models dominated by the cosmological
constant ($\Omega_\Lambda \sim 0.8$).

Empty regions as large as the Bo\"{o}tes void could have been
formed only from model 26 with $3\sigma$ perturbations (for a mass
perturbation of $10^{15}{\rm M}_\odot$). The diameter of the
baryonic component, $D_{\rm BF}=62h^{-1}{\rm Mpc}$, and the depth
obtained from our simulations ($\delta_{\rm BF}=-0.76$) are
similar to those determined by Dey et al.(1990) for
that large void.

Other CDM spectra studied here cannot account for the large voids
found by El-Ad et al.(1996) and M\"{u}ller et al. (2000) in their studies.
In particular, models 14, 20 and 23 could not account for the large voids even
considering $(2-3) \sigma$ negative density perturbations. The formation of the
largest voids could come, for these spectra, only from mergers of
two or more small voids as addressed, for example, by Dubinski
et al. (1993).

Our models show that the dark matter void is greater than the
baryonic matter void, this result being produced by the physical
processes acting on the baryons. We find a difference between the
radii of $\sim 1-10$ per cent, depending on the power spectra and
masses of the perturbations, when all physical processes are
included. This effect corresponds to the creation of a transition
zone defined by the radii of the baryonic and dark components. We
note that disregarding each one of the physical processes, the
thickness of this transition zone decreases and it goes to zero
when all processes and the pressure of the Universe are removed.
The analysis of the physical processes lead us to conclude that
the principal mechanism that acts on the evolution of voids
is photon drag. When we disregard this mechanism the final
diameter of the baryonic matter increases and the thickness of the
transition zone becomes almost half of its initial value.

Although the thickness of the transition zone is small (in the
range of $\sim 0.03-2.5h^{-1}{\rm Mpc}$) when compared to the
radii of the baryonic and dark components, interesting effects on
the formation of galaxies could occur. Putative objects formed
near voids and within the transition region would have a dark
matter to baryonic matter ratio such that $\rho_{\rm d}/\rho_{\rm
b} \neq \Omega_{\rm d} /\Omega_{\rm b}$. This means that the
amount of dark matter is different in such regions as compared to
that present in the Universe as a whole.

Then, if one uses these galaxies to determine by dynamical effects
or other techniques the quantity of dark matter present in the
Universe, the result obtained would be only local and not
representative of the Universe as a whole. Note, however, that
within voids the relation between baryonic matter and dark matter
maintains the same value as that of the Universe.

For different spectra one would obtain different results as
compared to those present here, but void regions with diameters of
tens of megaparsecs and transition zones of non-negligible sizes
would occur as well.

Owing to the fact that the baryonic matter undergoes a series of
physical processes, such as those considered here, that the
non-baryonic dark matter does not undergo, motions of matter on
large scale due to the presence of a gravitational potential could
induce a segregation between the two components, producing regions
with different values of $\rho_{\rm b} /\rho_{\rm d}$ that are
different from the $\Omega_{\rm b}/\Omega_{\rm d}$ relation of the
Universe.

It is important to note that the voids studied here are formed
only by the evolution of negative density perturbations. The voids
formed in this scenario are, therefore, cold. We do not consider a
possible energy injection, or excess of ionization (see, e.g.,
Srianand 1997), by quasars or other primordial objects near the
voids. In this way, the temperatures of the matter in our model
are never greater than the temperature of the radiation. The
inclusion of an energy source, such as quasars, is an interesting
possibility for a future extension of the scenario presented here.

Certainly the process of formation of the structures of the
Universe is a very complicated one. As discussed above, the
formation of voids and clusters of galaxies is part of the same
process of clustering. The formation of regions with positive
density contrasts, like galaxies and clusters of galaxies,
contributes to the growth of underdense regions. On the other hand,
the underdense regions push the matter around them during their
evolution and growth, contributing to the clustering processes.

In our study we only consider the evolution of negative density
perturbations, which is a small part of the very complicated
structure formation process. However, the assessment of the role
of a series of physical processes not considered in the
literature, besides including the presence of non-baryonic dark
matter and cosmological constant, are the main contributions of
the present study.

Also, because our study shows that the physical
processes considered here are relevant to the evolution of
negative density perturbations, this leads us to conclude that in
the structure formation of the Universe as a whole these very
processes could be relevant as well. Therefore, it would be of
interest to take into account, in more realistic modelling of the
structure formation, the various processes considered here.

\section*{Acknowledgments}
ODM and JCNdA would like to thank the Brazilian agency FAPESP
(grants 98/13735-7 and 00/00116-9, and 97/06024-4 and 97/13720-7,
respectively) and the Brazilian project Pronex/FINEP (no.
41.96.0908.00) for support. We would also like to thank the referee
for helpful comments that we feel considerably improved the paper.

\label{lastpage}


\begin{thebibliography}{99}
\bibitem{b1}  Aarseth S.J., Saslaw W.C., 1982, ApJ, 258, L7
\bibitem{b2}  Bahcall N., Soneira R., 1983, ApJ, 270, 20
\bibitem{b3}  Bertschinger E., 1985, ApJS, 58, 1
\bibitem{b4}  Carlberg R.G., 1981, MNRAS, 197, 1021
\bibitem{b5}  Cen R., 1998a, ApJ, 509, 16
\bibitem{b6}  Cen R., 1998b, ApJ, 509, 494
\bibitem{b7}  Chincarini G., Rood H.J., Thompson L.A., 1981, ApJ,
249, L47
\bibitem{b8}  Coles P., Lucchin F., 1995, Cosmology - The Origin and
Evolution of Cosmic Structure. Wiley, Chichester
\bibitem{b9}  da Costa L.N., Pellegrini P.S, Sargent W.L.W., Tonry J.,
Davis M., Meiksin A., Latham D.W., 1988, ApJ, 327, 544
\bibitem{b10}  de Araujo J.C.N., Opher R., 1988, MNRAS, 231, 923
\bibitem{b11}  de Araujo J.C.N., Opher R., 1989, MNRAS, 239, 371
\bibitem{b12}  de Araujo J.C.N., Opher R., 1990, ApJ, 350, 502
\bibitem{b13}  de Araujo J.C.N., Opher R., 1991, ApJ, 379, 461
\bibitem{b14}  de Araujo J.C.N., Opher R., 1993, ApJ, 403, 26
\bibitem{b15}  de Araujo J.C.N., Opher R., 1997, ApJ, 490, 488
\bibitem{b16}  Defouw R.J., 1970, ApJ, 161, 55
\bibitem{b17}  Dekel A., Silk J., 1986, ApJ, 303, 39
\bibitem{b18}  de Lapparent V., Geller M.J., Huchra J.P., 1986,
ApJ, 302, L1
\bibitem{b19}  Dey A., Strauss M.A, Huchra J.P., 1990, AJ, 99, 463
\bibitem{b20}  Dubinski J., da Costa L.N., Goldwirth D.S., Lecar
M., Piran T., 1993, ApJ, 410, 458
\bibitem{b21}  El-Ad H., Piran T., 1997, ApJ, 491, 421
\bibitem{b22}  El-Ad H., Piran T., da Costa L.N., 1996, ApJ, 462, L13
\bibitem{b23}  Filmore J.A., Goldreich P., 1984, ApJ, 281, 9
\bibitem{b24}  Geller M.J., Huchra J.P., 1989, Sci, 246, 897
\bibitem{b25}  Giovanelli R., Haynes M., Chincarini G., 1986, ApJ, 300, 77
\bibitem{b26}  Goldwirth D.S., da Costa L.N., van de Weygaert R.,
1995, MNRAS, 275, 1185
\bibitem{b27}  Gregory S.A., Thompson L.A., 1978, ApJ, 222, 784
\bibitem{b28}  Gregory S.A., Thompson L.A., Tifft W., 1981, ApJ, 243, 411
\bibitem{b29}  Hausman M.A., Olson D.W., Roth B.D., 1983, ApJ, 270, 351
\bibitem{b30}  Hoffman Y., Shaham J., 1983, in Abell G.O., Chincarini G.
eds, Proc. IAU Symp. 104, Early Evolution of the Universe and its Present
Structure. 203 Reidel, Dordrecht
\bibitem{b31}  Huchra J.P., Henry J.P., Postman M., Geller M.J., 1990,
ApJ, 365, 66
\bibitem{b32}  Ikeuchi S., Tomisaka K., Ostriker J.P., 1983, ApJ, 265, 583
\bibitem{b33}  Kirshner R.P., Oemler A., Jr., Schechter P.L.,
Schechtman S.A., 1981, ApJ, 248, L57
\bibitem{b34}  Kolb E.W., Turner M.S., 1990, The Early Universe. Addison
Wesley, New York
\bibitem{b35}  Lindner U., Einasto J., Einasto M., Freudling W., Fricke K.J.,
Tago E., 1995, A\&A, 301, 329
\bibitem{b36}  Lindner U., Fricke K.J., Einasto J., Einasto M., 1999,
in Katsuhiko S. ed, Proc. IAU Symp. 183, Cosmological Parameters and the
Evolution of the Universe. 185 Kluwer Academic, Dordrecht
\bibitem{b37}  Miranda O.D., Opher R., 1996, MNRAS, 283, 912
\bibitem{b38}  Miranda O.D., Opher R., 1997, ApJ, 482, 573
\bibitem{b39}  M\"{u}ller V., Arbabi-Bidgoli S., Einasto J.,
Tucker D., 2000, MNRAS, 318, 280
\bibitem{b40}  Oliveira S.R., Miranda O.D., de Araujo J.C.N., Opher R.,
1998a, MNRAS, 301, 101
\bibitem{b41}  Oliveira S.R., Miranda O.D., de Araujo J.C.N., Opher R.,
1998b, MNRAS, 301, 115
\bibitem{b42}  Opher R., Pires N., de Araujo J.C.N., 1997, MNRAS, 285, 811
\bibitem{b43}  Padmanabhan T., 1993, Structure Formation in the Universe.
Cambridge Univ. Press, Cambridge
\bibitem{b44}  Peebles P.J.E., 1968, ApJ, 153, 1
\bibitem{b45}  Peebles P.J.E., 1982, ApJ, 257, 438
\bibitem{b46}  Peebles P.J.E., 1993, Principles of Physical Cosmology.
Princeton Univ. Press, Princeton
\bibitem{b47}  Popescu C.C., Hopp U., Els\"{a}sser H., 1997, A\&A, 325,
881
\bibitem{b48}  Primack J.R., 1997, in Klapdor-Kleingrothaus H.V., Ramachers
Y. eds, Dark Matter in Astro and Particle Physics. World Scientific,
Singapore, 127
\bibitem{b49}  Richtmyer R.D., Morton K.W., 1967, Difference
Methods for Initial Value Problems. Interscience, New York
\bibitem{b50} Schmidt J.D., Ryden B.S., Melott A.L., 2001, ApJ, 546, 609
\bibitem{b51} Schwarz J., McCray R., Stein R.F., 1972, ApJ, 175, 673
\bibitem{b52}  Srianand R., 1997, ApJ, 478, 511
\bibitem{b53}  Szomoru A., van Gorkom J.H., Gregg M.D., Strauss M.A.,
1996a, AJ, 111, 2141
\bibitem{b54}  Szomoru A., van Gorkom J.H., Gregg M.D., Strauss M.A.,
1996b, AJ, 111, 2150
\bibitem{b55}  Thuan T.X., Gott J.R, Schneider S.E., 1987, ApJ, 315,
L93
\bibitem{b56}  Vettolani G., de Souza R.E., Marano B., Chincarini G.,
1985, A\&A, 144, 506
\end{thebibliography}
\end{document}